\documentclass[12pt,indent]{article}
\usepackage{a4,latexsym,amsmath,amssymb}
\usepackage[latin1]{inputenc}
\usepackage{float}
\usepackage{natbib}
\usepackage{graphics}
\usepackage[english]{babel}
\usepackage{tabularx}
\usepackage{lscape}
\usepackage{multirow}
\usepackage{rotating}
\usepackage{dsfont}
\usepackage{subfig}
\usepackage{bbm}
\usepackage[table]{xcolor}
\usepackage{tabu}
\usepackage{booktabs}
\usepackage{calc}
\usepackage{ifthen}
\usepackage{url}
\usepackage{colortbl}      
\newcommand{\node}{\operatorname{node}}
\newenvironment{tabularsmall}
{ \footnotesize \sffamily \tabular } {
\endtabular
\normalfont }

\newcommand{\betab}{{\boldsymbol{\beta}}}

\newcommand{\xb}{\boldsymbol{x}}

\newcommand{\zb}{\boldsymbol{z}}

\newcommand{\blanco}[1]{}

\def\d{\displaystyle}

\usepackage{natbib}
\usepackage{geometry}
\usepackage{setspace}

\begin{document}
\bibliographystyle{chicago}
\sloppy

\makeatletter
\renewcommand{\section}{\@startsection{section}{1}{\z@}%
        {-3.5ex \@plus -1ex \@minus -.2ex}%
        {1.5ex \@plus.2ex}%
        {\reset@font\Large\sffamily}}
\renewcommand{\subsection}{\@startsection{subsection}{1}{\z@}%
        {-3.25ex \@plus -1ex \@minus -.2ex}%
        {1.1ex \@plus.2ex}%
        {\reset@font\large\sffamily\flushleft}}
\renewcommand{\subsubsection}{\@startsection{subsubsection}{1}{\z@}%
        {-3.25ex \@plus -1ex \@minus -.2ex}%
        {1.1ex \@plus.2ex}%
        {\reset@font\normalsize\sffamily\flushleft}}
\makeatother



\newsavebox{\tempbox}
\newlength{\linelength}
\setlength{\linelength}{\linewidth-10mm} \makeatletter
\renewcommand{\@makecaption}[2]
{
  \renewcommand{\baselinestretch}{1.1} \normalsize\small
  \vspace{5mm}
  \sbox{\tempbox}{#1: #2}
  \ifthenelse{\lengthtest{\wd\tempbox>\linelength}}
  {\noindent\hspace*{4mm}\parbox{\linewidth-10mm}{\sc#1: \sl#2\par}}
  {\begin{center}\sc#1: \sl#2\par\end{center}}
}



\def\R{\mathchoice{ \hbox{${\rm I}\!{\rm R}$} }
                   { \hbox{${\rm I}\!{\rm R}$} }
                   { \hbox{$ \scriptstyle  {\rm I}\!{\rm R}$} }
                   { \hbox{$ \scriptscriptstyle  {\rm I}\!{\rm R}$} }  }

\def\N{\mathchoice{ \hbox{${\rm I}\!{\rm N}$} }
                   { \hbox{${\rm I}\!{\rm N}$} }
                   { \hbox{$ \scriptstyle  {\rm I}\!{\rm N}$} }
                   { \hbox{$ \scriptscriptstyle  {\rm I}\!{\rm N}$} }  }

\def\d{\displaystyle}

\title{Tree-Structured Modelling of Varying Coefficients}
\author{Moritz Berger, Gerhard Tutz \& Matthias Schmid}

\maketitle

\begin{abstract} 
\noindent
The varying-coefficient model is a strong tool for the modelling of interactions in generalized regression. It is easy to apply if both the variables that are modified as well as the effect modifiers are known. However, in general one has a set of explanatory variables and it is unknown which variables are modified by which covariates. A recursive partitioning strategy is proposed that is able to deal with  the complex selection problem. The tree-structured modelling yields for each covariate, which is modified by other variables, a tree that visualizes the modified effects. The performance of the method is investigated in simulations and two applications illustrate its usefulness. 
\end{abstract}

\noindent{\bf Keywords:} Varying-coefficient models; Interactions; Recursive partitioning; Tree-based models 

\section{Introduction} \label{sec:intro}

The generalized linear model is an established tool, which  has been widely applied  in regression problems. However, it is 
rather restricted when  interactions are needed. The inclusion of classical interaction terms in the form $x_jx_k\beta_{jk}$ still assumes a
 very rigid form of   interactions. More severe, they become awkward to interpret for higher order interaction terms that include more than 
two variables, in particular if they are on different scales. 

An alternative concept of interactions are effect modifiers or varying-coefficient models that were first introduced by \citet{HasTib:93}. The estimation of varying-coefficient models has been studied extensively in the literature. \citet{Fan2008statistical} give an comprehensive review on the varying-coefficient model and discuss several estimation approaches. 
Here, we consider alternative strategies how to set up regression models including varying coefficients defined by one or several effect modifiers. In particular we propose a tree-based strategy to determine which variables act as effect modifiers. In most applications of varying-coefficient models the effect modifiers are assumed to be known. However, in practice they are typically unknown and have to be selected from the pool of available variables. The proposed tree-structured approach (Section \ref{subsec:tree}) enables to simultaneously detect predictors 
with varying coefficients and the corresponding effect modifiers.

A tree-based approach to model varying coefficients using the traditional CART algorithm was proposed by \citet{Suetal:2011}. They use their method to asses the effect of an intervention program in a longitudinal breast cancer study. In the used model only the corresponding treatment effect was modified by other explanatory variables. Therefore, the result is exactly one tree. 
More recently, \citet{burgin2015tree} also proposed a tree-based model for varying coefficients. 
Their approach is similar in spirit to ours, but there are two crucial differences. First, the set of predictors $\xb$ and the set of effect modifiers $\zb$ (called moderators) are different and therefore have to be specified beforehand. Second, the model is explicitly designed for longitudinal studies only. \citet{wang2014boosted} use a boosted tree-based varying coefficient model with pre-specified effect modifiers in an application on product pricing.

Further related literature on varying-coefficient models refers mainly on regularization methods for the selection of smooth effect modifiers (see also Section \ref{subsec:smooth}). \citet{Wang2008} and \citet{Zhao2009} proposed to use the smoothly clipped absolute deviation penalty, \citet{leng2009simple} use a penalized likelihood method for smoothing spline ANOVA models and \citet{wang2009shrinkage} combine local polynomial smoothing and LASSO. \citet{Wong2009} moreover deal with concepts for missing data. 

The rest of the article is organized as follows: In Section \ref{sec:varying} we introduce the basic varying-coefficient model and the extended tree-structured model. Section \ref{sec:ill} contains an illustrative example based on artificial data. Details on the fitting procedure are given in Section \ref{sec:fitting}. In Section \ref{sec:sim} we show the results of several simulations examining the performance of the proposed model. Finally, in Section \ref{sec:app} the tree-structured model is illustrated by means of two applications.

\section{Varying-Coefficient Models} \label{sec:varying}
Let the data be given by $(y_i,\xb_{i}),\; i=1,\hdots,n$, where $y_i$ denotes the response and $\xb_i=(x_{i1},\hdots,x_{ip})^\top$ is a covariate vector of length $p$. In generalized linear models it is assumed that  the response $y_i$ given $\xb_{i}$  follows a simple exponential family. In addition, it is assumed that  the mean response $\mu_i=E(y_i|\xb_i)$ is linked to the explanatory variables by a  link function $g(\cdot)$ in the form $g(\mu_i)=\eta_i$, where $\eta_i=\xb_i^T\betab$ is a linear predictor. In the following modelling approaches the linear predictor is replaced by a more flexible predictor $\eta_i$ including varying coefficients.

\subsection{Smooth and Categorical Effect Modifiers} \label{subsec:smooth}
In varying-coefficient models, as proposed by \citet{HasTib:93} the predictor has the general form
\[
\eta_i=\beta_0+x_{i1}\beta_1(z_{i1})+x_{i2}\beta_2(z_{i2})+\hdots+x_{ip}\beta_p(z_{ip}),
\]
where $z_{i1},\hdots,z_{ip}$ denote additional predictors. It is assumed that $z_{i1},\hdots,z_{ip}$ change the coefficients of the predictors $x_{i1},\dots,x_{ip}$ through unspecified functions $\beta_1(\cdot),\dots, \beta_p(\cdot)$. Thus the $x$-variables have a linear effect, but the effects are modified by the so-called {\it effect modifiers} $z_{i1},\dots,z_{ip}$.

The effect modifiers can be continuous or categorical variables. If $z_{ij}$ is continuous it is typically assumed that $\beta_p(z_{ij})$ is a smooth function of unspecified form. Several strategies have been proposed for the estimation of these smooth functions, for example, by penalization, localization or boosting methods, see \citet{Hoover1998}, \citet{fan1999statistical}, \citet{Lu2008}, \citet{Wu1998}, \citet{KauTut:2000b} and \citet{Hofner2013variable}.
If the effect modifier is a categorical variable $z_{ij} \in \{1,\hdots,K\}$ the functions $\beta_j(z_{ij})$ are step functions of the form  $\sum_{k=1}^{K}{\beta_{jk} I(z_{ip}=k)}$, with indicator function $I(\cdot)$ and parameters $\beta_{j1},\hdots,\beta_{jK}$ yielding the predictor component
\[
x_{ij}\beta_1(z_{ij})=\sum_{k=1}^{K}x_{ij}{\beta_{jk} I(z_{ij}=k)}.
\]
Thus the coefficient on $x_{ij}$ depends on the value of $z_{ij}$. Since for categorical effect modifiers the number of parameters can become very large, tailored estimation strategies are needed for this kind of models. Penalization techniques for the estimation of models with categorical effect modifiers were proposed by \citet{GerTutEffMod:2012} for classical linear models and extended to generalized linear models by \citet{OelkerVarCoefs14}. In the categorical case one can additionally identify clusters of categories that share the same effect.

All the traditional approaches have in common that they aim at distinguishing between varying and non-varying coefficients. Given a specific effect modifier $z_{ij}$ one wants to know if the effects of $x_{ij}$ is constant over the whole range of $z_{ij}$ or varies across values of $z_{ij}$. Thus,  typically it is assumed that one knows which variable is a potential effect modifier. Then one determines the way it modifies coefficients.

\subsection{Tree-Structured Modelling of Varying Coefficients} \label{subsec:tree}

The models and estimation strategies described in the previous section have some limitations and drawbacks. If one uses these models 
the effect modifier has to be specified beforehand. However, usually it is   totally unclear which variable should be considered as a 
relevant effect modifier. Moreover, it is not known if a varying coefficient is determined by just one variable or if more than one 
effect modifier determine the varying coefficient. It is even possible that varying coefficients are caused by the interaction of several 
effect modifiers. The  recursive partitioning method proposed in the following provides a  solution to these problems. By recursive splitting the method itself 
identifies the effect modifiers that induce varying coefficients if they are present.

Since we do not assume that the effect modifiers are known, we consider only the set of covariates $x_1,\hdots,x_{p}$. If effect modifiers are present they are from this set and modify coefficients on covariates from this set. A simple example with just one effect modifier $x_{ij}$ is  given by the predictor 
\[
\eta_i=\beta_0+x_{i1}\beta_1(x_{ij})+ \hdots+x_{ij}\beta_{j}(x_{ij})+ \hdots+x_{i,p}\beta_{p}(x_{ij}),
\] 
where we define $\beta_{j}(x_{ij})=\beta_{j}$. That means, if the effect modifier is identical to the covariate that is modified by it, it is fixed. Therefore, the predictor is 
\begin{equation} \label{mod:treeone}
\eta_i=\beta_0+x_{i1}\beta_1(x_{ij})+ \hdots+x_{ij}\beta_{j}+ \hdots+x_{i,p}\beta_{p}(x_{ij}).
\end{equation} 
In particular, we assume that the predictor $\eta_i$ contains a main effect of $x_{ij}$, if $x_{ij}$ modifies the other variables.
The predictor specified by equation \eqref{mod:treeone} is understood as a generic form of the predictor. If a variable is categorical with $K$ values (on a nominal scale) the corresponding variable $x_{ij}$ typically contains $K-1$ dummy variables.

The  principle of recursive partitioning or tree-based modelling is that the predictor space is recursively split into a set of rectangles. Within each rectangle a simple model (for example, a constant) is fitted. The most popular version goes back to \citet{BreiFrieOls:84} and is known as \textit{classification and regression trees} (CART).  
Trees for varying-coefficients work in the same way. However, the splits refer to the coefficients. Therefore, successively one chooses a coefficient (corresponding to a predictor), a variable (the effect modifier)  and a split point to split the coefficient into two disjoint regions. In each region the coefficient is then fitted by a constant. 

\subsubsection*{Models with Three Covariates}

For simplicity we first consider three covariates $x_1$, $x_2$ and $x_3$ that are metrically scaled or ordinal.
Suppose that $x_2$ is an effect modifier that changes the effect of $x_1$. Then a split in the coefficient of the covariate $x_1$ generated by the effect modifier  $x_2$ at split point $c_2$ means to fit a model with predictor
\[
\eta_i=\beta_0+x_{i1}\left[\beta_{1\ell}^{[1]}I(x_{i2}\leq c_2)+\beta_{1r}^{[1]}I(x_{i2}> c_2)\right]+x_{i2}\beta_2+x_{i3}\beta_3,
\]
where $I(\cdot)$ denotes the indicator function with $I(a)=1$ if $a$ is true and $I(a)=0$ otherwise. The parameter $\beta_{1\ell}^{[1]}$ denotes the effect of $x_1$ in the (left) region $\{x_{i2}\leq c_2\}$ and  $\beta_{1r}^{[1]}$ denotes the effect of $x_1$ in the (right) region $\{x_{i2}> c_2\}$. If $x_2$ is a binary covariate, like gender, one obtains the two effects
\[
\beta_{1\ell}^{[1]}=\beta_{1,male} \text{ for males} \quad \text{ and } \quad \beta_{1r}^{[1]}=\beta_{1,female} \text{ for females.}
\]
If the effects of covariate $x_1$ are further modified by $x_3$, a second split (for example in the left node) with regard to $x_3$ and split point $c_3$ yields the two daughter nodes
\[
I(x_{i2}\leq c_2)I(x_{i3}\leq c_3) \text{ and } I(x_{i2}\leq c_2)I(x_{i3}> c_3),
\]
and the model with predictor
\begin{align*}
\eta_i=\beta_0+x_{i1}&\left[\beta_{1\ell}^{[2]}I(x_{i2}\leq c_2)I(x_{i3}\leq c_3)+\beta_{1r}^{[2]}I(x_{i2}\leq c_2)I(x_{i3}> c_3)\right.\\
&\left.+\beta_{1r}^{[1]}I(x_{i2}> c_2)\right ]+x_{i2}\beta_2+x_{i3}\beta_3,
\end{align*}
where $\beta_{1\ell}^{[2]}$ and $\beta_{1r}^{[2]}$ are the new effects in the regions built by the second split. After further splits in the coefficients of covariate $x_1$ regarding $x_2$ and $x_3$ the resulting model has the form
\begin{equation} \label{mod:tree_x1}
\eta_i=\beta_0+x_{i1}tr_{1}(x_{i2},x_{i3})+x_{i2}\beta_2+x_{i3}\beta_3,
\end{equation}
where $tr_{1}(x_{i2},x_{i3})=\sum_{q=1}^{Q}{\beta_{1q}\node_{1q}(x_{i2},x_{i3})}$  represents   a tree determined by splits in $x_2$ and $x_3$. Each node is defined by a product of several indicator functions of the form
\[
\node(x_{i2},x_{i3})=\prod_{b=1}^{B} I(x_{ij_b} > c_{j_b})^{a_{b}}I(x_{ij_b} \leq c_{j_b})^{1-a_{b}},
\]
where $B$ is the total number of branches, $c_{jb}$ is the selected split point in variable $x_{j_b} \in \{x_2,x_3\}$ and $a_b \in \{0,1\}$ indicates which of the two indicator functions is involved. The terminal nodes contain the varying coefficients $\beta_{1q}$ of covariate $x_1$.

The model specified by equation \eqref{mod:tree_x1} contains the two main effects of the effect modifiers $x_2$ and $x_3$. If $x_2$ and $x_3$ (or one of the two) have an effect on the response, the relation can be simply linear or again varying over the other two variables, respectively. More specifically, the effect of $x_2$ can be modified by $x_1$ and $x_3$ and the effect of $x_3$ can be modified by $x_1$ and $x_2$. Hence, the tree-structured model with three covariates has the form
\[
\eta_i=\beta_0+x_{i1}tr_{1}(x_{i2},x_{i3})+x_{i2}tr_{2}(x_{i1},x_{i3})+x_{i3}tr_{3}(x_{i1},x_{i2}),
\]
which is composed of three potential trees defined by splits in at most two variables, respectively.

\subsubsection*{The General Model with p Covariates}

In the general case one has $p$ covariates $x_j,\; j \in \{1,\hdots,p\}$ and the coefficients of each one can be modified by all the other variables $x_m,\;m \in \{1,\hdots,p\}\setminus j$. The tree component of the model regarding $x_j$ is then given by
\[
tr_{j}(x_{im})=\beta_{j\ell}^{[1]}I(x_{im}\leq c_m)+\beta_{jr}^{[1]}I(x_{im}> c_m).
\]
To determine the first split for a fixed covariate $x_j$, means to select the best model among all possible effect modifiers $x_m$ and corresponding split points $c_m$. This corresponds to examine all the null hypotheses $H_0:\beta_{j\ell}^{[1]}=\beta_{jr}^{[1]}$. If $H_0$ cannot be rejected for any combination of effect modifier and split point the covariate is considered to have a linear effect on the response. To examine the null hypotheses, likelihood ratio (LR) tests are used in our procedure. In the very first step one chooses the combination of coefficient, effect modifier and split point with the smallest $p$-value. If a significant effect is found, the first split is carried out for the selected covariate. In Section \ref{sec:fitting} details are given on the splitting criterion.

The second split is either in the coefficients of the same or another covariate, with regard to the same or another effect modifier. As is in all later steps the search is the same, but for predictors that have already been split one starts from already built nodes. If a predictor is never selected for splitting it is assumed to have the simple linear effect $\beta_j$. The procedure stops if no significant effect is found anymore  (see Section \ref{sec:fitting}).

After termination of the algorithm, let $V \subseteq \{x_1,\hdots,x_p\}$ denote the subset of covariates that have been selected for splitting, $L \subseteq \{x_1,\hdots,x_p\}\setminus V$ denote the subset of covariates with a linear effect on the response (not selected for splitting) and $M_j \subseteq \{x_1,\hdots,x_p\}\setminus x_j$ the subset of effect modifiers for covariate $x_j$. For covariates that have never been selected for splitting $M_j$ is an empty set. If no split is performed at all, that is the simple linear model without varying coefficients is valid, $V$ is an empty set. On the other hand, if the coefficients of all covariates are modified by at least one other variable, $L$ is an empty set. In the extreme case, without any influential covariate (pure intercept model), both $V$ and $L$ are empty. 

Using this notation, the tree-structured model in its most general form can be written by
\begin{equation}\label{eq:model}
\eta_i=\beta_0+\sum_{x_j \in V} x_{ij} tr_{j}(M_j)+\sum_{x_{\ell} \in L} x_{i\ell}\beta_{\ell}.
\end{equation}
The method yields an individual tree for each covariate that shows varying coefficients. If varying coefficients are present the relevant effect modifiers are selected simultaneously.
In the last step of the proposed algorithm (described in Section 4) the linear effects of covariates that were not chosen for splitting during iteration and do not serve as effect modifier for any other covariate are tested for significance. Nevertheless, to follow the hierarchical principle, a linear effect of a variable that was chosen as effect modifier will remain in the model. Therefore, it is ensured that the model includes a (linear or non linear) main effect of each effect modifier.   

In the following we use the abbreviation TSVC for tree-structured varying coefficient model. 

\section{An Illustrative Example} \label{sec:ill}

\begin{figure}[t]
\centering
\includegraphics[width=0.49\textwidth]{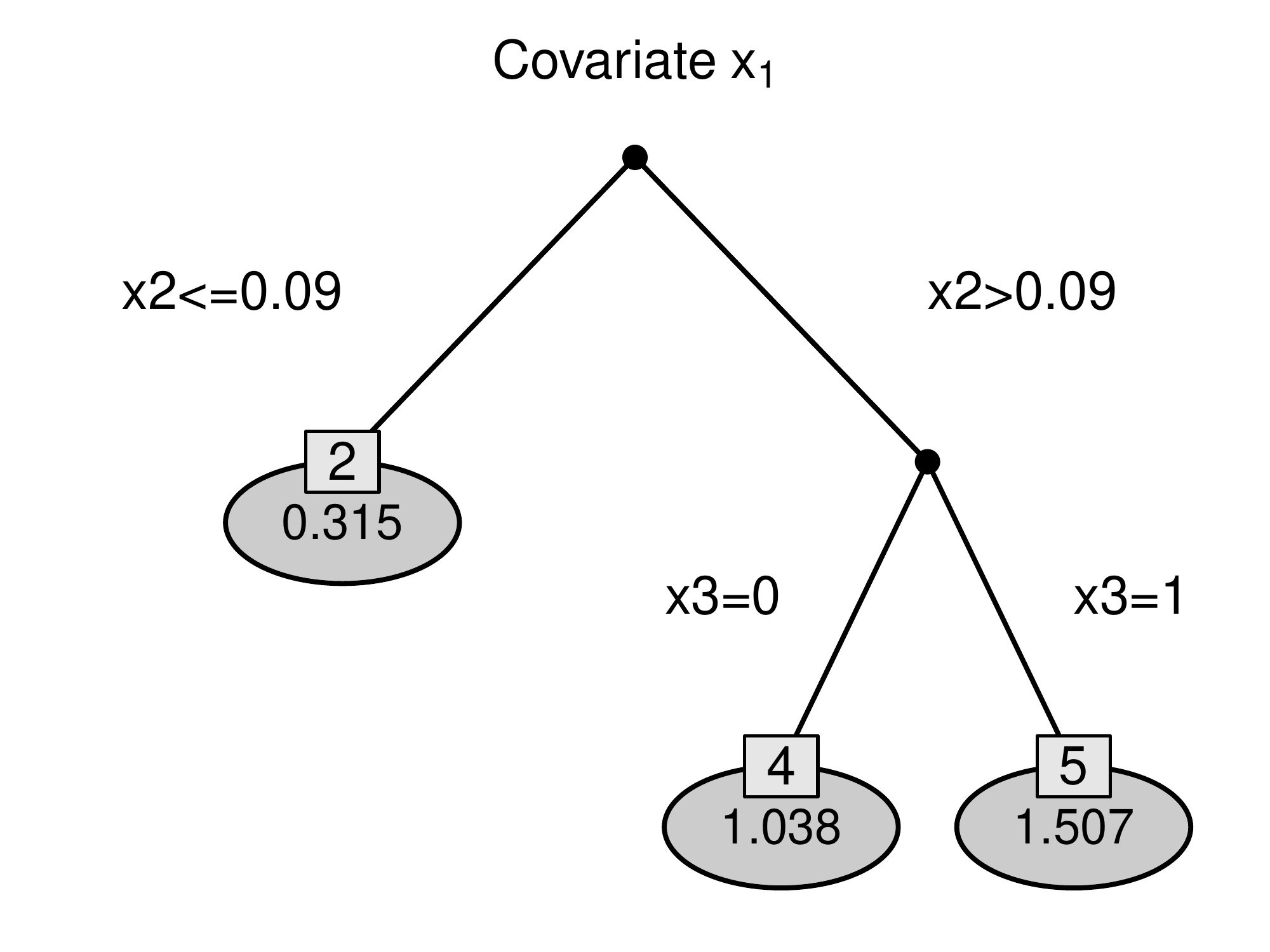}
\includegraphics[width=0.49\textwidth]{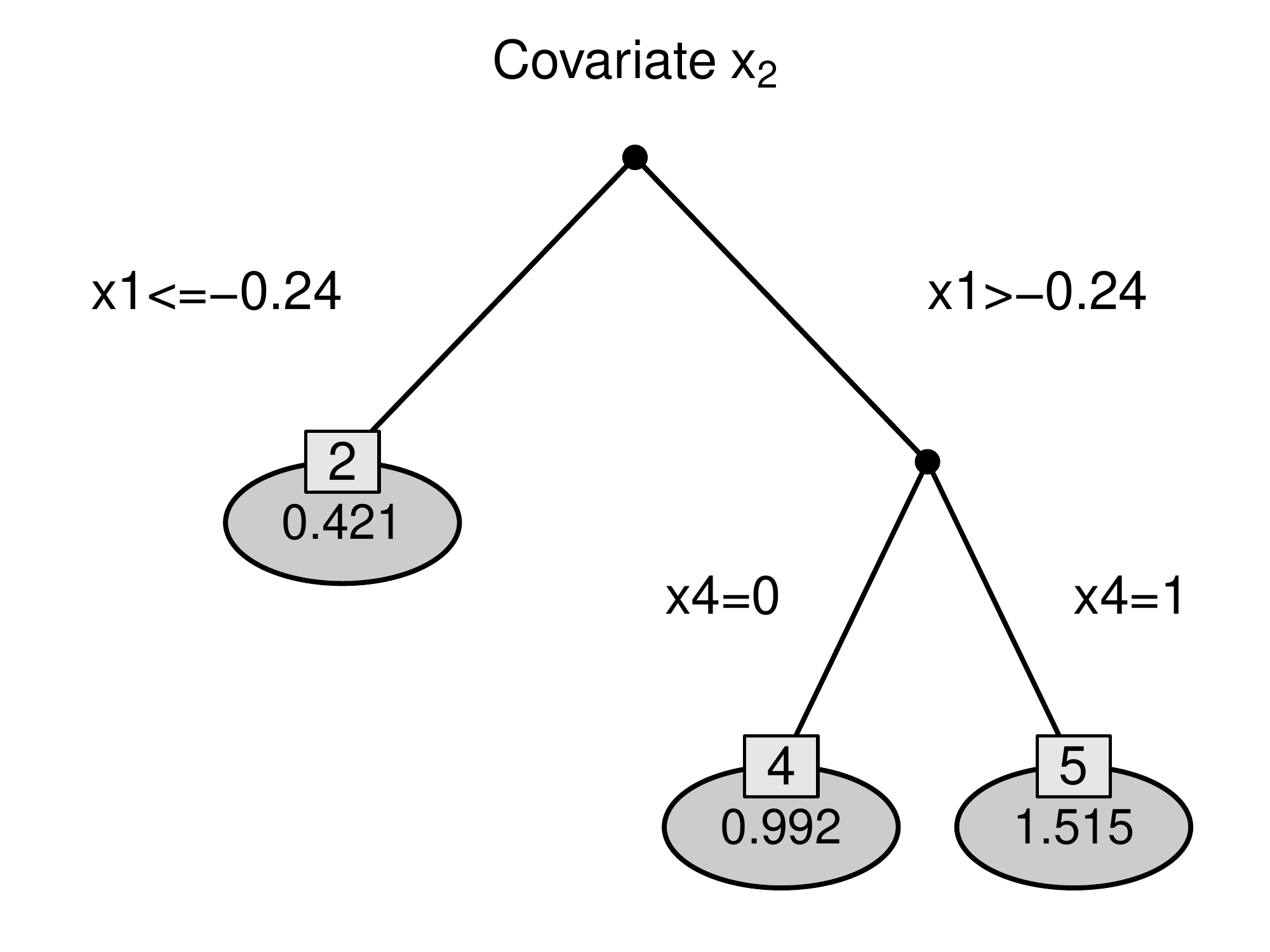}
\caption{Estimated trees for covariate $x_1$ and $x_2$ of the illustrative example. The estimated coefficients are given in each leaf of the trees.}
\label{fig:trees_illustrative}
\end{figure}

To illustrate the proposed TSVC we make use of artificial data. We consider data with normally distributed response $y_i\sim N(\mu_i,1), i=1,\hdots, 400$. The linear predictor of the model is composed of two continuous covariates $x_{i1},x_{i2} \sim N(0,1)$ and two binary covariates $x_{i3},x_{i4} \sim B(1,0.5)$. The predictor of the model is given by $\eta_i=\beta_0+x_{i1}tr_{1}(x_{i2},x_{i3})+x_{i2}tr_{2}(x_{i1},x_{i4})+x_{i3}\beta_3+x_{i4}\beta_4$.
It is assumed that the effects of $x_3$ and $x_4$ on the response are simply linear. But, the effect of $x_1$ is modified by $x_2$ and $x_3$ and is determined by the tree component
\[
tr_{1}(x_{i2},x_{i3})=\beta_1+0.6I(x_{i2}>0.2)+0.6I(x_{i2}>0.2 \cap x_{i3}=1).
\]
The effect of $x_2$ is modified by $x_1$ and $x_4$ and determined by
\[
tr_{2}(x_{i1},x_{i4})=\beta_2+0.6I(x_{i1}> -0.2)+0.6I(x_{i1}> -0.2 \cap x_{i4}=1).
\]
The true coefficients are $\beta_0=0.2$ and $\beta_1=\beta_2=\beta_3=\beta_4=0.4$. Figure \ref{fig:trees_illustrative} shows the resulting trees for 
one exemplary estimation. In this example the true underlying tree structure is detected for both covariates and no further split is performed in the coefficients of any other covariate. The estimates of the linear terms are $\hat{\beta}_0=0.249$, $\hat{\beta}_3=0.331$ and $\hat{\beta}_4=0.461$ and therefore close to 
the true values. It is seen from the trees that there are three groups represented by three terminal nodes, respectively. For covariate $x_1$ it is distinguished between $\{x_2 \leq 0.09\}$ and $\{x_2 > 0.09\}$, and within this group between $\{x_3=0\}$ and $\{x_3=1\}$. Together with the estimates given in the leafs of tree, these results are exactly in line with the true simulated effects. This also holds for covariate $x_2$, see right part of Figure \ref{fig:trees_illustrative}. Due to the data generating process the simulated split points regarding the continuous covariates $x_2$ and $x_1$ must not necessarily be present in the data, but the detected ones are very close to them.

\section{Fitting Procedure} \label{sec:fitting}

In this section we give a detailed description of the algorithm that yields the proposed TSVC.

\subsection{Concepts}\label{subsec:concepts}

When building trees, the most important parameter is the number of splits that determines the depth and hence the size of the trees. 
There are several strategies to determine the adequate size of the trees. In traditional approaches one typically grows large trees 
and prunes them to an adequate size afterward, see \citet{BreiFrieOls:84} and \citet{Ripley:96}. An alternative strategy, which is applied here, 
is to directly control the size of the trees by early stopping. In each step of the tree growing one decides if a further split is needed or not. For an introduction into the basic concept of so-called \textit{conditional inference trees}, see \citet{Hotetal:2006}.

In each step of the algorithm one selects the best split among all the predictors, effect modifiers and corresponding split points. 
This is done by examining all the null hypotheses $H_0:\beta_{j\ell}=\beta_{jr}$ against the alternatives $H_1:\beta_{j\ell}\neq\beta_{jr}$ and 
by choosing the combination with the smallest $p$-value of the corresponding LR-test. A simple criterion is to stop splitting when the $p$-value 
exceeds a pre-specified threshold $\alpha$. However, when $\alpha$ is intended to have meaningful interpretation as a global type I error level 
(see below), one should adapt for multiple testing errors because in each split a huge number of hypotheses is tested. In the presence of many 
covariates and a potentially large number of splits, procedures like the Bonferroni correction will lead to local significance levels 
$\alpha$ close to zero and hence will be not suited for tree construction. Therefore, we apply a concept based on maximally selected statistics. 
The idea is to investigate the dependence of the response and the selected effect modifier at a global level that  
takes the number of splits into account.

Let us focus on one predictor $x_j$ and one effect modifier $x_m$ with possible split points $c_m$. When a split point is selected based on the 
LR test statistic $T_{m,c_m}$ one investigates the distribution of $T_m = max_{c_m}{T_{m,c_m}}$. Typically the test statistics $T_{m,c_m}$ are 
strongly correlated. The $p$-value that can be obtained by the distribution of $T_m$ provides a measure for the relevance of effect modifier $x_m$. The result is not influenced by the number of split points, see \citet{HotLau:03}, \citet{Shih:04}, \citet{ShiTsa:2004}. During tree building splitting is stopped when the global hypothesis of independence between the response and the selected effect modifier cannot be rejected. Similar to the unified framework proposed by \citet{Hotetal:2006}, the method explicitly accounts for the
involved multiple testing problem and provides an unbiased recursive partitioning scheme that avoids the selection bias toward variables with many possible splits. 
To calculate the asymptotic distribution under the null hypothesis of $T_m$ and to derive a test decision a permutation test is used. 
That means one permutes the values of effect modifier $x_m$ in the original data and computes the corresponding value of the test statistic. For a huge number of permutations one obtains an approximation of the distribution under the null hypothesis
and a corresponding $p$-value. To determine the $p$-values with sufficient accuracy, the number of permutations should increase with the number of variables (potential effect modifiers) in the model.

\subsubsection*{Meaning of the Threshold $\alpha$}

Given overall significance level $\alpha$, the local significance level $\alpha_{\ell}$ for one permutation test is set to $\alpha_{\ell}=\alpha/(p-1)$, where $p-1$ corresponds to the number of potential effect modifiers. With this adaption the probability to falsely identify varying coefficients for each predictor is controlled by $\alpha$. If one has $N$ predictors without varying coefficients, one can expect $N\alpha$ predictors to be falsely selected for splitting. Using $\alpha_{\ell}=\alpha/(p-1)$ at the same time ensures that on the level of the predictor the family-wise error rate is under control. That means, the probability of falsely identifying at least one effect modifier (for fixed predictor) is controlled by $\alpha$. 

\subsection{Basic Algorithm}

\hrule
\vspace{0.5cm}
\begin{center}{\bf Basic Algorithm}\end{center}

\begin{description}
\item{\it Step 1 (Initialization)}

Set counter $\nu=1$

\begin{itemize}
\item[(a)] Estimation

For all covariates $x_j,\, j=1,\hdots,p$, fit all the candidate models with predictor
\begin{align*}
\eta_i=&\beta_0+x_{ij}\left(\beta_{j1}I(x_{im}\leq c_{mk})+\beta_{j2}I(x_{im}> c_{mk})\right)+\sum_{\ell \in \{1,\hdots,p\}\setminus j}{x_{i\ell}\beta_{\ell}},\\
&m=\{1,\hdots,p\}\setminus j,\quad k=1,\hdots,K_m.
\end{align*}

\item[(b)] Selection

Select the model that has the best fit. Let $c_{m_1,k_1}$ denote the best split, which is found for covariate $x_{j_1}$ and effect modifier $x_{m_1}$.

\item[(c)] Splitting Decision

Select the predictor and effect modifier with the largest value of $T_m$. Carry out permutation test for this combination with significance level $\alpha_{\ell}$. If significant, fit the selected model yielding estimates $\hat{\beta_0}$, $\hat{\beta}_{j_1,1}$, $\hat{\beta}_{j_1,2}$ and $\hat{\betab}_{\ell}$, and nodes $node_{j_1,1}, node_{j_1,2}$, $\nu=2$. If not, stop, no varying coefficients detected.
\end{itemize}

\item{\it Step 2 (Iteration)}

\begin{itemize}
\item[(a)] Estimation

For all covariates $x_j,\, j=1,\hdots,p$, and already built nodes $q=1,\hdots,Q_{j\nu}$, fit all the candidate models with new coefficients (while the rest of the model remains the same)
\[
\beta_{j,Q_{j\nu}+1}node_{jq}I(x_{im}\leq c_{mk})+\beta_{j,Q_{j\nu}+2}node_{jq}I(x_{im}>c_{mk})
\]
for all $m=\{1,\hdots,p\}\setminus j$ and remaining, possible split points $c_{mk}$.

\item[(b)] Selection

Select the model that has the best fit yielding the split point $c_{m_\nu,k_\nu}$, which is found for covariate $x_{j_\nu}$ in node $node_{j_\nu,q_\nu}$ and effect modifier $x_{m_\nu}$

\item[(c)] Splitting Decision

Select the node and effect modifier with the largest value of $T_m$. Carry out permutation test for this combination with significance level $\alpha_{\ell}$. If significant, fit the selected model yielding the additional estimates $\hat{\beta}_{j_\nu,Q_{j_\nu,\nu}+1}, \hat{\beta}_{j_\nu,Q_{j_\nu,\nu}+2}$, set $\nu=\nu+1$. If not, stop.

\end{itemize}

\item{\it Step 3 (Linear Term)}

Collect the selected covariates $x_{j_\nu}$ in $V$ and set $L=\{x_1,\hdots,x_p\}\setminus V$. Collect the effect modifiers for the $j$-th covariate in $M_j$ and set $M = \bigcup M_j$. 

\begin{itemize}

\item[(a)] Selection

For all covariates $x_{\ell} \in L \setminus M$, examine the null hypothesis $H_0:\beta_{\ell}=0$, by use of a permutation test with significance level $\alpha$. If not significant, exclude $x_{\ell}$ from $L$.

\item[(b)] Estimation

Estimate final model with components $\hat{\beta}_0$, $\hat{tr}_j(M_j)$ and $\hat{\betab}_{\ell}$, where $M_j$ comprises all effect modifiers $x_{m_\nu}$ for which $x_{j_\nu}=j$ holds.

\end{itemize}

\end{description}
\hrule


\section{Numerical Experiments} \label{sec:sim}

In the following we investigate the performance of the proposed TSVC. We are in particular interested in the ability of the procedure to detect the predictors with varying coefficients and the corresponding effect modifiers. In all simulation scenarios the responses $y_i,\;i=1,\hdots,n$, are normally distributed with noise variable $\epsilon_i \sim N(0,\sigma_{\epsilon}^2)$. The models include two standard normally distributed covariates, $x_1,\; x_2 \sim N(0,1)$ and two binary covariates, $x_3,\; x_4 \sim B(1,0.5)$. We consider scenarios with $n \in \{100, 250, 500\}$ observations and standard deviation $\sigma_{\epsilon}  \in \{1, 1.5, 2\}$. In each setting 100 data
sets were generated. During estimation each permutation test was based on 1000
permutations.

\subsubsection*{Evaluation Criteria}

In order to evaluate the proposed model \eqref{eq:model} we compute true positive rates (TPR) and false positive rates (FPR). We distinguish between TPR and FPR on the covariate level and for the combination of covariate and effect modifier. Let $\delta_j,\, j=1,\hdots,p$, be the indicator, with $\delta_j=1$ if covariate $x_j$ exhibits varying coefficients induced by any effect modifier and $\delta_j=0$ otherwise. In addition, let $\delta_{jm}$ be the indicator, with $\delta_{jm}=1$ if covariate $x_j$ exhibits varying coefficients with regard to effect modifier $x_m,\; m=\{1,\hdots,p\}\setminus j$ and $\delta_{jm}=0$ otherwise. With indicator function $I(\cdot)$, criteria to judge the identification of varying coefficients are: 
\begin{itemize}
\item[-] True positive rate on the covariate level: 

$TPR_C=\frac{1}{\#\{j:\delta_j=1\}}\sum_{j:\delta_j=1}{I(\hat{\delta}_j=1)}$

\item[-] False positive rate on the covariate level: 

$FPR_C=\frac{1}{\#\{j: \delta_j=0\}}\sum_{j:\delta_j=0}{I(\hat{\delta}_j=1)}$
\item[-] True positive rate for the combination of covariate and effect modifier:
 
$TPR_{CM}=\frac{1}{\#\{j,m:\delta_{jm}=1\}}\sum_{j,m:\delta_{jm}=1}{I(\hat{\delta}_{jm}=1)}$

\item[-] False positive rate for the combination of covariate and effect modifier: 

$FPR_{CM}=\frac{1}{\#\{j,m:\delta_{jm}= 0\}}\sum_{j,m:\delta_{jm}= 0}{I(\hat{\delta}_{jm}=1)}$

\end{itemize}

\subsection{Models without Varying Coefficients}

We start with simulations, where the data generating model is a simple linear model, that is no varying coefficients are 
present (\textit{scenario 1}). The true model has the form $\mu_i=\beta_0+x_{i1}\beta_1+x_{i2}\beta_2+x_{i3}\beta_3+x_{i4}\beta_4$, with coefficients 
$\beta_0=0.2$ and $\beta_1=\beta_2=\beta_3=\beta_4=0.4$. This results in a simulated $R^2$ of $0.28$ ($\sigma_{\epsilon}=1$), $0.15$ 
($\sigma_{\epsilon}=1.5$) and $0.09$ ($\sigma_{\epsilon}=2$). The absence of varying coefficients is a baseline situation to check a 
possible inflation of the false positive rate on the covariate level, which is explicitly controlled for by the significance level $\alpha$ in the algorithm (see Section \ref{subsec:concepts}). Table \ref{tab:sim1_V} shows the false positive rates on the covariate level ($FPR_C$) for the nine settings with varying error variance and sample size, as the average over the 100 repetitions, respectively. It is seen, that the significance level is kept in all settings. For a small sample size ($n=200$) the approach is even conservative. Conspicuously the results do not differ with the error variance. 

In addition, the proportion of covariates (PoC) that were included in the model (either with a main effect or split in a tree) is given in Table \ref{tab:sim1_power}. The values correspond to the average probabilities over the 100 repetitions, respectively. With four predictors $x_1, \hdots,x_4$, the value $0.750$ means, that on average one covariate (with a true main effect) was excluded from the model. Except for the setting with $\sigma_{\epsilon}=2$ and $n=100$ $(R^2=0.09)$ the proportion of covariates in the model is quite high $(> 0.5)$, which indicates a good performance of the TSVC.

\begin{table}[!t]
\centering
\begin{tabularsmall}{lccccccccc}
\toprule
\bf{Scenario 1}&\multicolumn{3}{c}{$\sigma_{\epsilon}=1$}&\multicolumn{3}{c}{$\sigma_{\epsilon}=1.5$}&\multicolumn{3}{c}{$\sigma_{\epsilon}=2$}\\
&n=100&n=250&n=500&n=100&n=250&n=500&n=100&n=250&n=500\\
\hline
$FPR_C$&0.020&0.045&0.040&0.020&0.045&0.040&0.020&0.045&0.040\\
\bottomrule
\end{tabularsmall}
\caption{Average false positive rates on the covariate level for simulation scenario 1 without varying coefficients.}
\label{tab:sim1_V}
\end{table}  

\begin{table}[!t]
\centering
\begin{tabularsmall}{lccccccccc}
\toprule
\bf{Scenario 1}&\multicolumn{3}{c}{$\sigma_{\epsilon}=1$}&\multicolumn{3}{c}{$\sigma_{\epsilon}=1.5$}&\multicolumn{3}{c}{$\sigma_{\epsilon}=2$}\\
&n=100&n=250&n=500&n=100&n=250&n=500&n=100&n=250&n=500\\
\hline
$PoC$&0.762&0.912&0.997&0.520&0.757&0.922&0.362&0.632&0.810\\
\bottomrule
\end{tabularsmall}
\caption{Average proportion of covariates in the model for simulation scenario 1 without varying coefficients.}
\label{tab:sim1_power}
\end{table}
  
\subsection{Models with Smooth Effect Modifier}

In a second simulation we consider data with smooth effect modifiers (\textit{scenario 2}). Here, the true, underlying model has the form $\mu_i=\beta_0+x_{i1}f_1(x_{i2})+x_{i2}f_2(x_{i1})+x_{i3}\beta_3+x_{i4}\beta_4$, with coefficients $\beta_0=0.2$ and $\beta_3=\beta_4=0.4$ and smooth functions 
\[
f_j(x)=arctan(x),\quad j=1,2.
\]
By definition, there are sigmoidal relations between $x_2$ and the regression coefficients of $x_1$ and between $x_1$ and the regression coefficients of $x_2$. The data generating process is determined by smooth functions, so that by successive splitting in one covariate the algorithm should be able to capture the underlying functional form. 

Average true positive rates and false positive rates on the covariate level (first and third row) as well as for the combination of covariate and effect modifier (second and fourth row) are given in Table \ref{tab:sim2_VM}. It is seen from the true positive rates that the method shows good overall performance: For all settings the $TPR_C$ and $TPR_{CM}$ are higher than $0.5$. It is noteworthy that the TPRs for the combination of covariate and effect modifier are almost exactly the same as those for covariates only. Therefore, if a significant split is found, it is always with regard to the right effect modifier. False positive rates are very small throughout all settings, in particular the global significance level (approximately) holds.

\begin{figure}[!t]
\centering 
\includegraphics[width=0.8\textwidth]{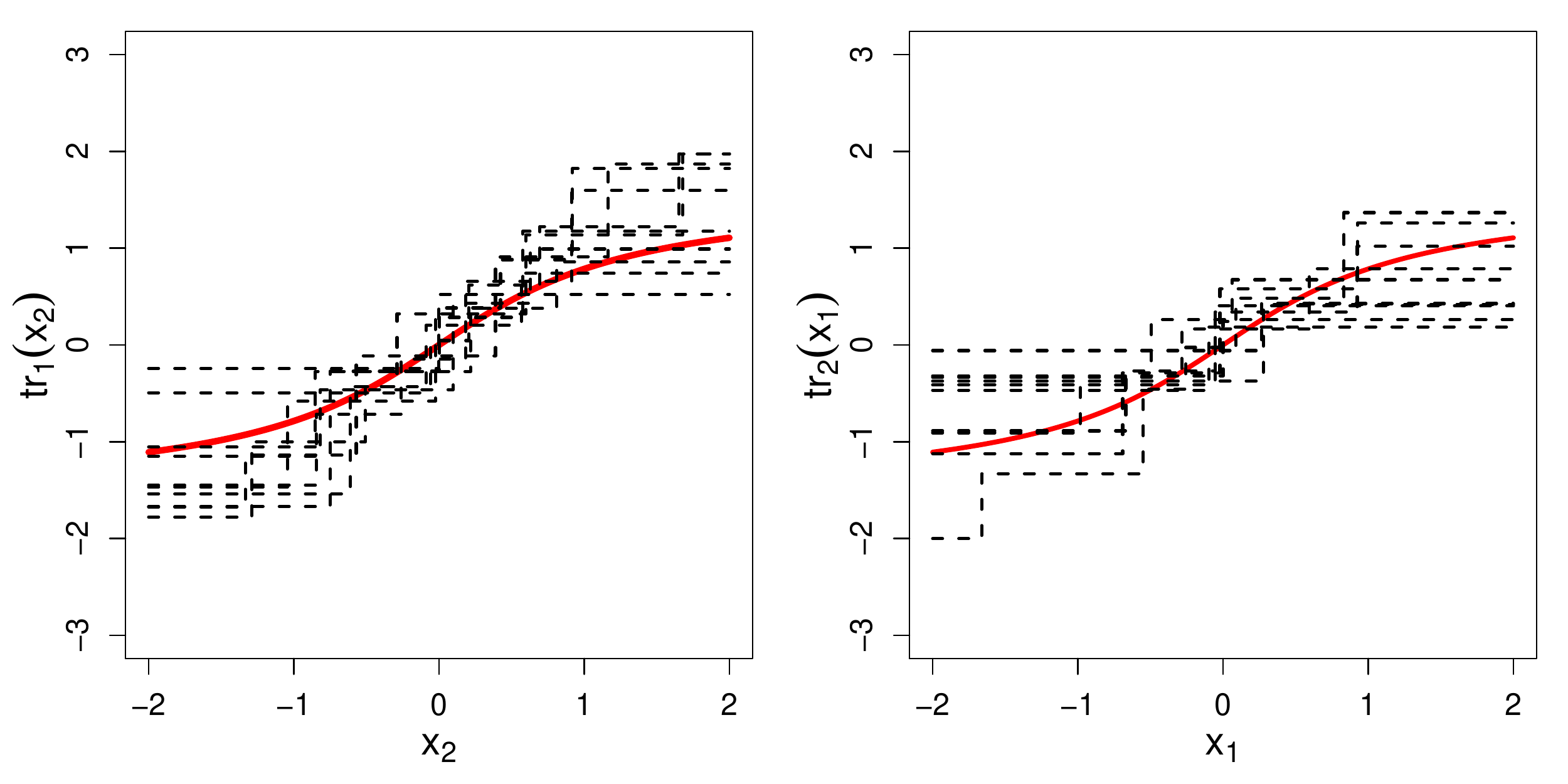}
\caption{Estimated step functions $\hat{tr}_1(x_2)$ and $\hat{tr}_2(x_1)$ for 10 randomly chosen replications of simulation scenario 2 with $\sigma=1$ and $n=500$. The true functions are drawn with solid lines.}
\label{fig:sim2_visu}
\end{figure}

Figure \ref{fig:sim2_visu} visualizes the true functions $f_1(x_2)$ and $f_2(x_1)$ (solid lines) and the estimated trees $\hat{tr}_1(x_2)$ and $\hat{tr}_2(x_1)$ (dashed lines) for 10 randomly chosen replications of the simulation with $\sigma=1$ and $n=500$ for the range $x1,\,x2 \in [-2,2]$. It is seen that in both cases the estimated step functions approximate the true smooth functions. 

\begin{table}[!t]
\centering
\begin{tabularsmall}{lccccccccc}
\toprule
\bf{Scenario 2}&\multicolumn{3}{c}{$\sigma_{\epsilon}=1$}&\multicolumn{3}{c}{$\sigma_{\epsilon}=1.5$}&\multicolumn{3}{c}{$\sigma_{\epsilon}=2$}\\
&n=100&n=250&n=500&n=100&n=250&n=500&n=100&n=250&n=500\\
\hline
$TPR_C$&0.785&0.900&0.970&0.635&0.790&0.920&0.555&0.680&0.855\\
$TPR_{CM}$&0.785&0.900&0.970&0.635&0.790&0.920&0.555&0.675&0.855\\
\cline{2-10}
$FPR_C$&0.070&0.060&0.030&0.055&0.055&0.040&0.045&0.050&0.020\\
$FPR_{CM}$&0.017&0.022&0.012&0.014&0.019&0.014&0.011&0.014&0.013\\
\bottomrule
\end{tabularsmall}
\caption{Average true positive rates and false positive rates for simulation scenario 2 with smooth effect modifiers.}
\label{tab:sim2_VM}
\end{table}  

\begin{table}[!t]
\centering
\begin{tabularsmall}{lccccccccc}
\toprule
&\multicolumn{3}{c}{$\sigma_{\epsilon}=1$}&\multicolumn{3}{c}{$\sigma_{\epsilon}=1.5$}&\multicolumn{3}{c}{$\sigma_{\epsilon}=2$}\\
&n=100&n=250&n=500&n=100&n=250&n=500&n=100&n=250&n=500\\
\hline
\bf{Scenario 3}\\
$TPR_C$&0.140&0.365&0.490&0.060&0.175&0.360&0.045&0.140&0.215\\
$TPR_{CM}$&0.115&0.335&0.470&0.040&0.130&0.330&0.025&0.085&0.175\\
\cline{2-10}
$FPR_C$&0.020&0.020&0.025&0.015&0.020&0.025&0.015&0.020&0.020\\
$FPR_{CM}$&0.009&0.018&0.020&0.007&0.018&0.016&0.007&0.017&0.014\\
\hline
\bf{Scenario 4}\\
$TPR_C$&0.115&0.295&0.465&0.055&0.160&0.285&0.045&0.100&0.170\\
$TPR_{CM}$&0.080&0.265&0.440&0.025&0.115&0.250&0.015&0.065&0.130\\
\cline{2-10}
$FPR_C$&0.026&0.041&0.0250&0.021&0.038&0.028&0.028&0.041&0.030\\
$FPR_{CM}$&0.004&0.006&0.004&0.003&0.006&0.005&0.004&0.006&0.005\\
\hline
\bf{Scenario 5}\\
$TPR_C$&0.315&0.495&0.535&0.135&0.350&0.500&0.075&0.250&0.425\\
$TPR_{CM}$&0.170&0.340&0.482&0.062&0.192&0.350&0.035&0.122&0.275\\
\cline{2-10}
$FPR_C$&0.025&0.030&0.035&0.020&0.020&0.030&0.025&0.020&0.030\\
$FPR_{CM}$&0.011&0.017&0.020&0.008&0.012&0.016&0.008&0.012&0.016\\
\bottomrule
\end{tabularsmall}
\caption{Average true positive rates and false positive rates for simulation scenarios 3, 4 and 5 with varying coefficients induced by discrete splits.}
\label{tab:sim345_VM}
\end{table} 

\subsection{Varying Coefficients Induced by Discrete Splits}

In this section we consider three scenarios with varying coefficients that are induced by splits with regard to one or more effect modifiers. First we assume the two binary covariates $x_1$ and $x_2$ as effect modifiers (\textit{scenario 3}) and simulate data from the model $\mu_i=\beta_0+x_{i1}\beta_1+x_{i2}\beta_2+x_{i3}tr_3(x_{i4})+x_{i4}tr_4(x_{i3})$, where $tr_3(x_{i4})=\beta_3+0.4 I(x_{i4}=0)$ and $tr_4(x_{i3})=\beta_4+0.4 I(x_{i3}=0)$. The regression coefficients $\beta_0,\hdots,\beta_4$ are set the same as in the previous simulations. Subsequently we increase the number of covariates in the model (\textit{scenario 4}). We add  $x_5,\; x_6 \sim N(0,1)$ and $x_7,\; x_8 \sim B(1,0.5)$. The four additional covariates are not influential with $\beta_5=\hdots=\beta_8=0$.  Hence the selection of covariates and effect modifiers might be more challenging. In a last simulation (\textit{scenario 5}) we consider a more complex structure, where the effects of $x_3$ and $x_4$ are additionally modified by $x_2$. The simulated trees have the form 
\begin{align*}
tr_3(x_{i4},x_{i2})&=\beta_3+0.4 I(x_{i4}=0)+0.4 I(x_{i4}=0 \cap x_{i2}>0) \quad \text{and}\\ 
tr_4(x_{i3},x_{i2})&=\beta_4+0.4 I(x_{i3}=0)+0.4 I(x_{i3}=0 \cap x_{i2}>0).
\end{align*}
An overview of all the results for the \textit{scenarios 3, 4 and 5} is given in Table \ref{tab:sim345_VM}. Each value again corresponds to the average over 100 replications. As the overall strength of the effects is weaker than in \textit{scenario 2}, the true positive rates are considerably smaller throughout all settings. The false positive rates on the covariate level are consistently very small, in particular the global significance level holds (with a tendency of the method to be conservative). It is also seen that, in particular in \textit{scenario 5}, the hit rates for the combination of covariate and effect modifier ($TPR_{CM}$) are a little smaller than the hit rates for covariates. Thus the algorithm is not always able to detect both effect modifiers.

\section{Applications} \label{sec:app}

In order to demonstrate the utility and the potential of the proposed TSVC we show the results of two real data examples. 

\subsection{Swiss Labour Market}

\begin{table}[t]
\caption{Summary statistics of the response (participation) and the covariates of the Swiss data (on the original scale).}
\begin{center}
\begin{tabularsmall}{lcccccc}
\toprule
\bf{covariate}&\multicolumn{6}{c}{\bf{summary statistics}}\\
\midrule
participation&&0: 471&&&1: 401&\\
\midrule
&$x_{min}$&$x_{0.25}$&$x_{med}$&$\bar{x}$&$x_{0.75}$&$x_{max}$
\\[0.2em]
income (\$)&1322&35320&41900&47730&53470&237000\\
age (years)&20&32&39&39.96&48&62\\
education&1&8&9&9.307&12&21\\
youngkids&0&0&0&0.311&0&3\\
oldkids&0&0&1&0.983&2&6\\
\\[0.4em]
foreign&&0: 656&&&1: 216&\\
\bottomrule
\end{tabularsmall}
\end{center}
\label{tab:swiss_desc}
\end{table}

\begin{table}[t]
\caption{Parameters estimates, standard errors and $z$ values of the simple logistic regression model for the Swiss data.}
\begin{center}
\begin{tabularsmall}{lrrr}
\toprule
\bf{covariate}&\bf{estimate}&\bf{std error}&\bf{z value}\\
\midrule
income&-0.815&0.205&-3.966\\ 
age      &   -0.510   & 0.090 & -5.638 \\
education   & 0.031  &  0.029 & 1.093  \\   
youngkids  &-1.330  &  0.180 & -7.386 \\
oldkids   &  -0.021  &  0.073 & -0.298 \\   
foreign    &  1.310 &   0.199 &  6.560 \\
\midrule
\midrule
deviance&&1052.8&\\
AIC&&1066.8&\\
\bottomrule
\end{tabularsmall}
\end{center}
\label{tab:swiss_linear}
\end{table}

\begin{table}[t]
\caption{Overview on the results of the proposed TSVC model in the modelling of the Swiss data.}
\begin{center}
\begin{tabularsmall}{lr}
\toprule
\bf{covariate}&\bf{estimate}\\
\midrule
income & $tr(age)$\\
age &  -1.042\\
education & ---\\
youngkids& $tr(age, foreign)$\\
oldkids& -0.230\\
foreign &  1.058 \\
\midrule
\midrule
deviance&1004.8\\
AIC&1022.8\\
\bottomrule
\end{tabularsmall}
\end{center}
\label{tab:swiss_tree}
\end{table}

\begin{figure}[t]
\begin{center}
\includegraphics[width=0.49\textwidth]{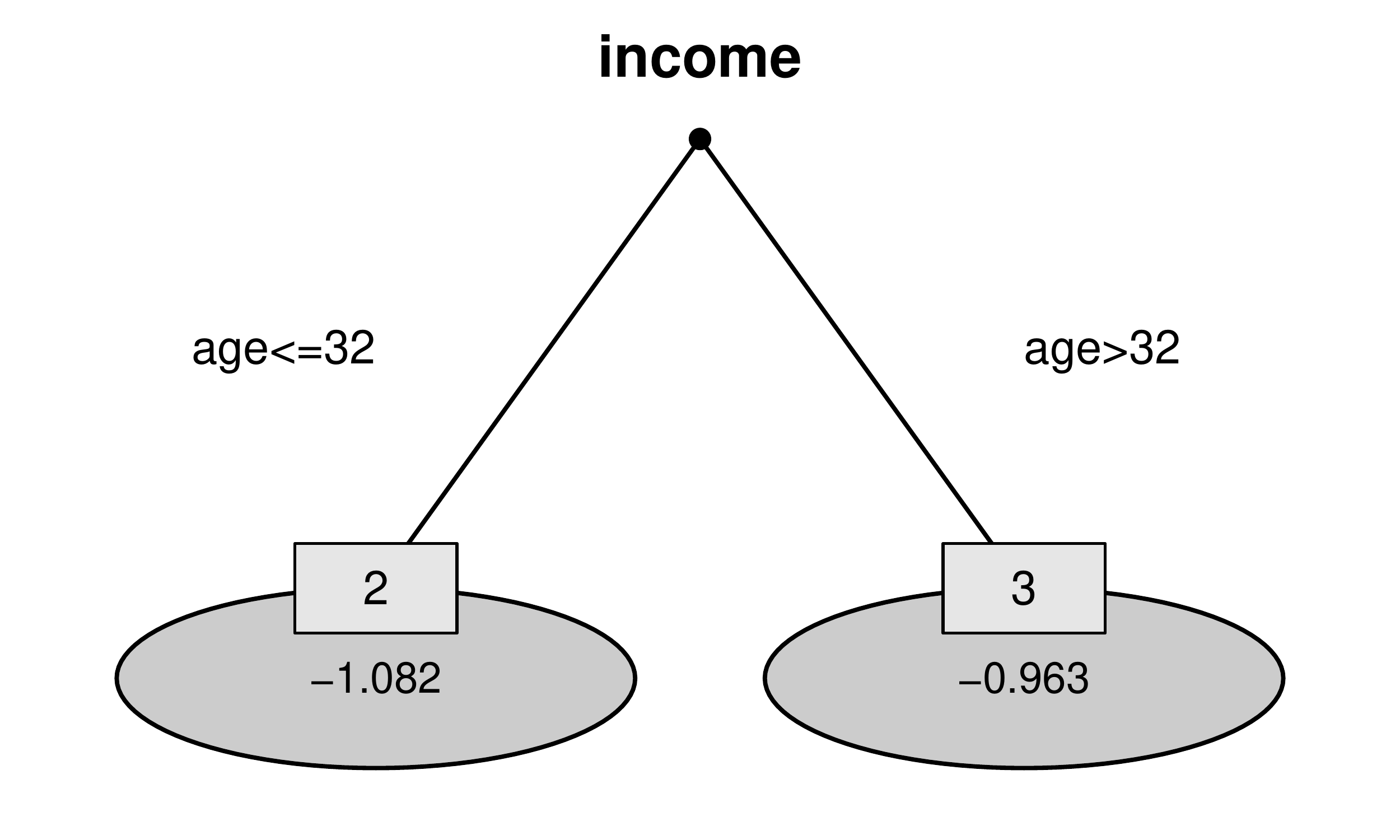}
\includegraphics[width=0.49\textwidth]{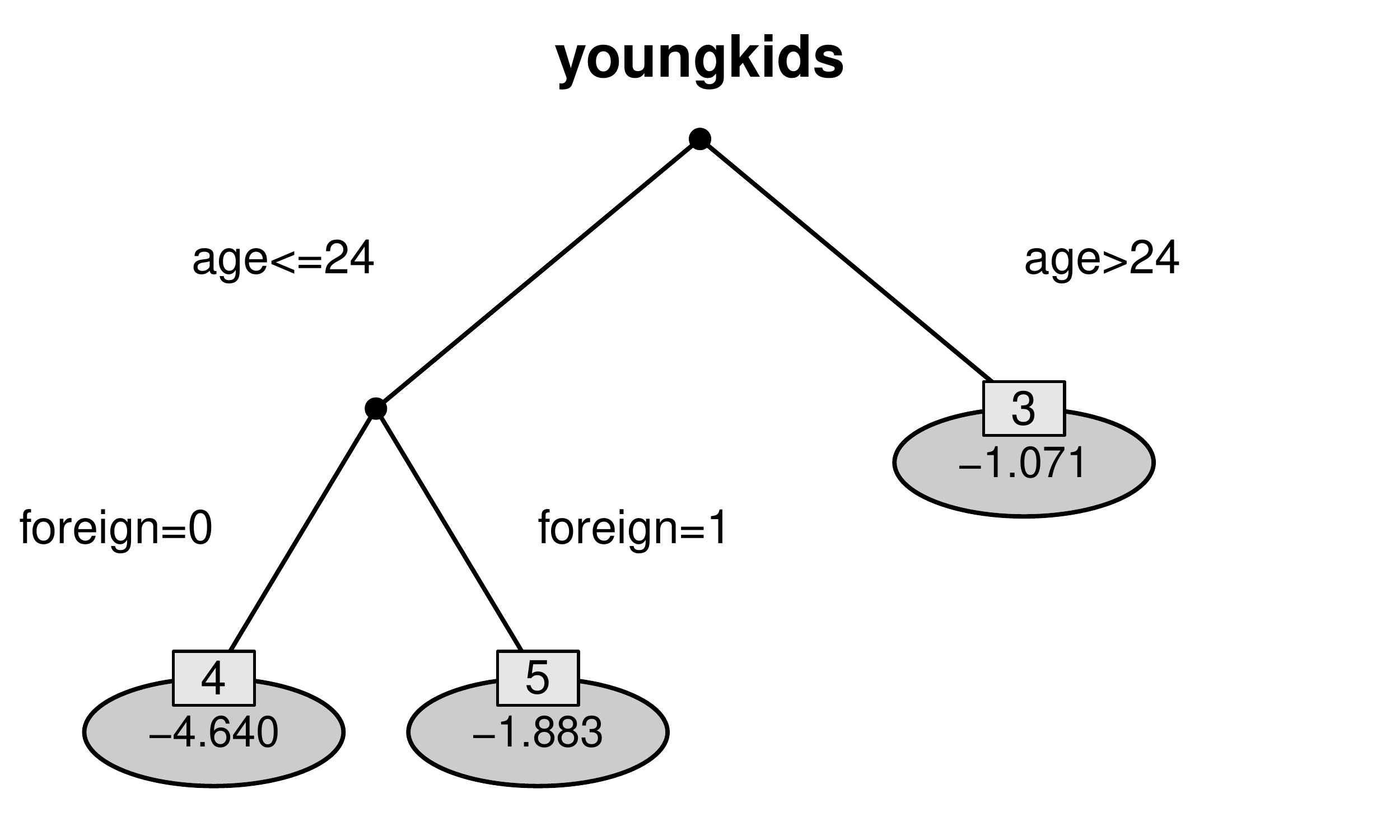}
\caption{Estimated trees by the TSVC model for covariates income and youngkids (Swiss data). The varying coefficients are given in the leafs of the trees, respectively.}
\label{fig:swiss_tree}
\end{center}
\end{figure}

We consider data from the health survey SOMIPOPS for Switzerland in 1981. The data set is available in the \texttt{R} package \texttt{AER} \citep{AER} and was analysed before by \citet{gerfin1996}. The data consists of a sample of 872 married women living in Switzerland. The response that is investigated is the binary outcome if the individual participates in the labour market ($y_i=1$) or not ($y_i=0$). The response is modelled by a logistic regression model of the form $logit(P(y_i=1|\xb_i))=\eta_i$. The explanatory variables that are included in the model are the logarithm of the yearly non-labour income, the age in decades (centered around 40), the years of formal education, the number of young children under 7 years of age, the number of older children over 7 years of age and an indicator if the individual is a foreigner (not Swiss). The summary statistics of the response and the original covariates are given in Table \ref{tab:swiss_desc}. With $401$ working and $471$ non-working women, the response is rather balanced.

The estimates, standard errors and $z$ values obtained by a simple logistic regression model are given in Table \ref{tab:swiss_linear}. 
It is seen that there are significant effects for all the covariates except for the education of the women and the number of children older than 
7 years of age (oldkids). The parameter estimates indicate that the chance to participate in the labour market decreases with increasing 
non-labour income, increasing age and with the number of young children. Interestingly, non-Swiss women ($foreign=1$) are more likely to participate in the 
labour market than Swiss women. 

When fitting the proposed TSVC model the results differ considerably (see the overview in Table \ref{tab:swiss_tree}). The algorithm performs 
three splits with regard to the coefficients of income and youngkids, until further splits are not significant ($\alpha=0.05$). In addition 
there are linear effects of age, foreign and oldkids. Hence, in contrast to the model with a linear predictor the TSVC indicates that the 
number of older children is also significantly associated with the response. The covariate education is completely excluded from the model.

The resulting trees for income and youngkids are shown in Figure \ref{fig:swiss_tree}. The varying coefficients of income are induced by 
age (left panel). Generally, non-labour income has a negative effect on the probability to participate in the labour market, but the effect 
is even stronger for younger women ($age<=32$). The varying coefficients in youngkids are induced by the effect modifiers age and foreign 
(right panel). As already seen from the model with a linear predictor, the chance to participate in the labour market decreases with 
the number of young children. However, the strength of the effects are much more differentiated, depending on the sub-group. 
The effect is strongest for comparably young women ($age<=24$) with Swiss citizenship ($foreign=0$), but is much attenuated for older women ($age>24$).     

For the TSVC model one obtains the residual deviance $1004.8$ and AIC $1022.8$, which are substantially smaller values than those for the model with linear predictor (compare Table \ref{tab:swiss_linear}). Therefore, the tree-structured model is to be preferred over the simple model.

\subsection{Australian Health Service Utilization}

\begin{table}[t]
\caption{Summary statistics of the response (visits) and the covariates of the AHS data (on the original scale).}
\begin{center}
\begin{tabularsmall}{lcccccc}
\toprule
\bf{covariate}&\multicolumn{6}{c}{\bf{summary statistics}}\\
\midrule
&$x_{min}$&$x_{0.25}$&$x_{med}$&$\bar{x}$&$x_{0.75}$&$x_{max}$\\
visits&0&0&0&0.302&0&9\\
\midrule
&$x_{min}$&$x_{0.25}$&$x_{med}$&$\bar{x}$&$x_{0.75}$&$x_{max}$\\
\\[0.2em]
income (\$)&0&2500&5500&5832&9000&15000\\
age (years)&19&22&32&40.640&62&72\\
illness&0&0&1&1.432&2&5\\
reduced&0&0&0&0.862&0&14\\
health&0&0&0&1.218&2&12\\
\\[0.4em]
gender&&0: 2488&&&1: 2702&\\
private&&0: 2892&&&1: 2298&\\
freepoor&&0: 4968&&&1: 222\hphantom{0}&\\
lchronic&&0: 4585&&&1: 605\hphantom{0}&\\
\bottomrule
\end{tabularsmall}
\end{center}
\label{tab:AHC_desc}
\end{table}

\begin{table}[t]
\caption{Parameters estimates, standard errors and $z$ values of the simple poisson regression model for the AHS data.}
\begin{center}
\begin{tabularsmall}{lrrr}
\toprule
\bf{covariate}&\bf{estimate}&\bf{std error}&\bf{z value}\\
\midrule
gender    &   0.170 &  0.055  & 3.055  \\
income    &  -0.199 &  0.084 & -2.364  \\
age       &   0.042  & 0.013  & 3.123  \\
illness   &   0.194 &  0.017 & 11.007  \\
reduced   &   0.126 &  0.005 & 25.180  \\
health    &   0.031 &  0.010 &  3.099  \\
private   &   0.087 &  0.053 &  1.626  \\
freepoor  &  -0.465 &  0.176 & -2.641  \\
lchronic  &   0.071 &  0.066 &  1.081  \\
\midrule
\midrule
deviance&&4384.3&\\
AIC&&6735.9&\\
\bottomrule
\end{tabularsmall}
\end{center}
\label{tab:AHC_linear}
\end{table}

\begin{table}[t]
\caption{Overview on the results of the proposed TSVC model in the modelling of the AHS data.}
\begin{center}
\begin{tabularsmall}{lr}
\toprule
\bf{covariate}&\bf{estimate}\\
\midrule
gender    & ---\\
income    &  $tr(reduced, illness)$\\
age       &  $tr(reduced)$\\
illness   &  0.111 \\
reduced   &  0.161 \\
health    &  0.042 \\
private   &  ---  \\
freepoor  &  -0.499 \\
lchronic & $tr(reduced)$\\
\midrule
\midrule
deviance&4089.9\\
AIC&6447.5\\
\bottomrule
\end{tabularsmall}
\end{center}
\label{tab:AHC_tree}
\end{table}

\begin{figure}[t]
\begin{center}
\includegraphics[width=0.49\textwidth]{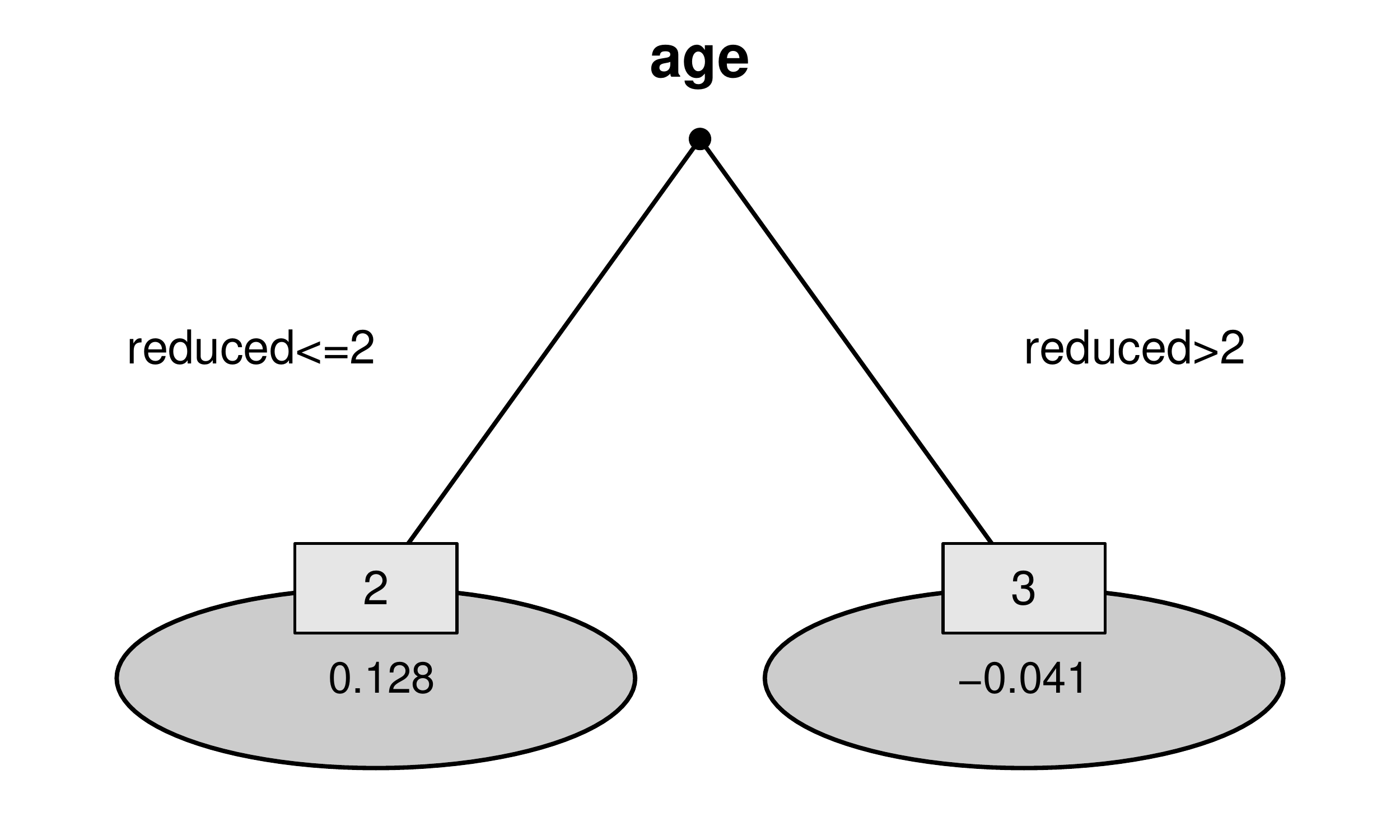}
\includegraphics[width=0.49\textwidth]{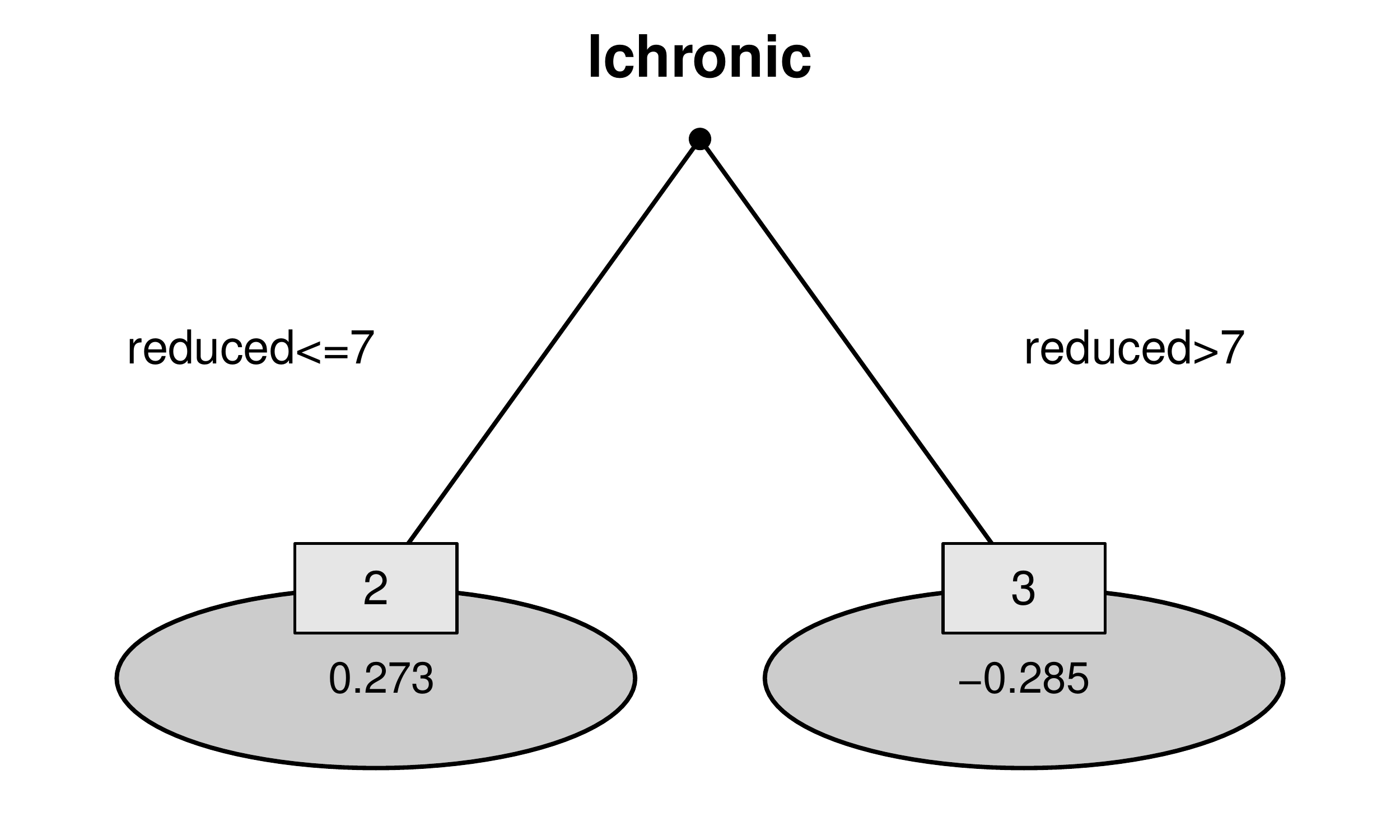}

\includegraphics[width=0.98\textwidth]{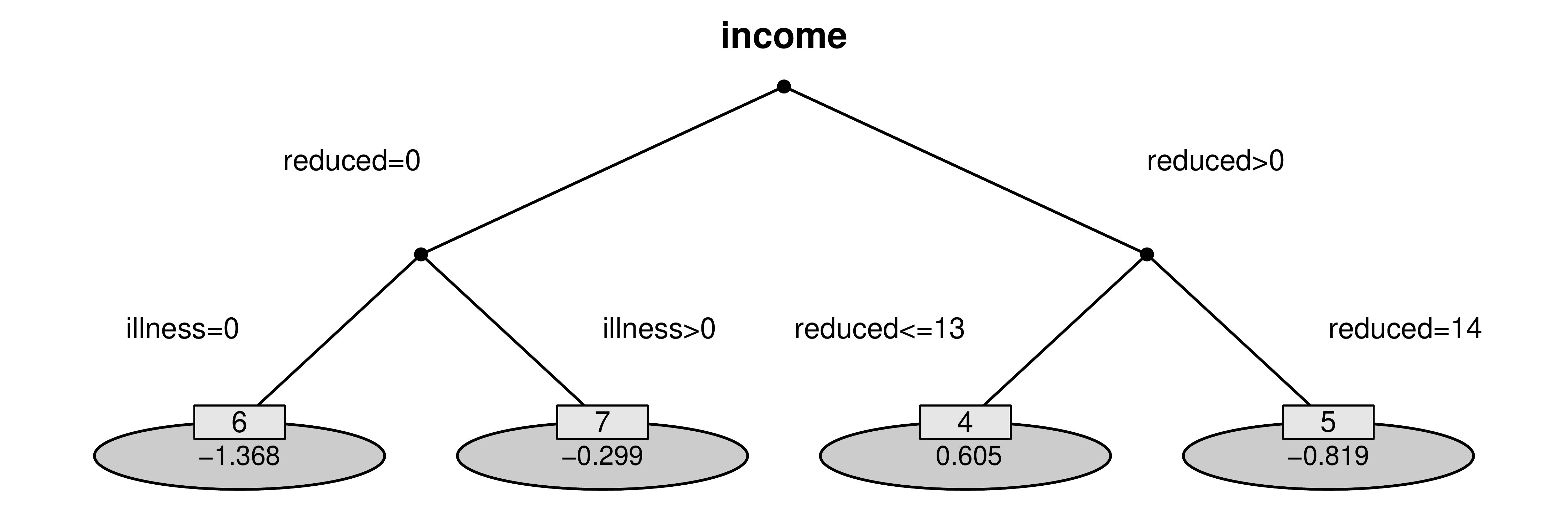}
\caption{Estimated trees by the TSVC model for covariates age, lchronic and income (AHS data). The varying coefficients are given in the leafs of the trees, respectively.}
\label{fig:trees_visits}
\end{center}
\end{figure}

In a second application we consider cross-section data originating from the $1977/1978$ Australian Health Survey (AHS). The original data 
set was used before by \citet{CamTri:86, CamTri:98} and is also available from the \texttt{R} package \texttt{AER} \citep{AER}.  The data 
set contains information on the number of visits to a doctor or specialist for 5190 individuals over 18 years of age in the two-week period 
before an interview (visits). The response is modeled by a Poisson model of the form $\mu_i=exp(\eta_i)$. Explanatory variables, which are 
used for modelling, are the gender, the income in tens of thousands of dollars, the age in decades (centered around 40), the number of 
illnesses in the past two weeks (illness), the number of days of reduced activity in the past two weeks due to illness or injury (reduced), 
the health questionnaire score by Goldberg's method (health) and the three indicators if the individual has a private health insurance (private), 
if the individual has free government health insurance due to low income (freepoor) as well as if there is a chronic condition limiting activity 
(lchronic). The dummy variable private represents a higher level of insurance cover, whereas freepoor represents just a basic insurance cover. 
Table \ref{tab:AHC_desc} shows a  descriptive overview on the response and the explanatory variables. About $79\%$ of the respondents had zero 
consultations, indicating overdispersion in the data. For simplicity, this fact will be omitted in the models discussed in the following.

The results when using a simple Poisson model with a linear predictor are given in Table \ref{tab:AHC_linear}. As expected, 
it is seen from the $z$ values that the number of illnesses and the number of days of reduced activity are strongest associated with 
the number of visits to a doctor. Respondents holding a free government health insurance (freepoor) are expected to visit the doctor less 
often. By contrast, there is no significant effect for the group of respondents holding a private health insurance. A chronic condition 
limiting activity also does not show a significant association with the response. 

An overview on the results of the proposed TSVC model is given in Table \ref{tab:AHC_tree}. It is seen that private and gender 
(which shows a significant effect in the simple model) are completely excluded from the tree-structured model. Influential covariates 
that are still in the model as linear effects are illness and reduced. They additionally serve as effect modifiers for the coefficients of 
 income and lchronic. Furthermore, there is a positive linear effect for health and again a strong negative effect for freepoor. 

The estimated trees are pictured in Figure \ref{fig:trees_visits}. The upper left tree for age shows, that the expected number of visits 
increases with the age of the respondents if the number of days of reduced activity is low ($reduced <= 2$), otherwise the effect does not appear. 
From the tree for lchronic in the upper right (which was not significant in the simple model)  it is seen, that a chronic condition increases the 
expected number of doctor's visits for all respondents with up to one week of reduced activity ($reduced <=7$). Interestingly, there is exactly the 
opposite effect if the period of reduced activity is longer. This difference is canceled out in the simple model (see Table \ref{tab:AHC_linear}) 
and is therefore not visible. 

Remarkable differences occur for the effect of income as seen from  the tree in the lower panel of Figure \ref{fig:trees_visits}. 
An increasing income strongly reduces the expected number of doctor's visits for all the respondents without reduced activity and without 
an illness ($reduced=0$, $illness=0$). The effect of income is also reduced if the period of reduced activity lasts the entire two weeks ($reduced=14$). The estimate 
$-1.368$ means a reduction of the expected mean by the factor $exp(-1.368)=0.255$. However, for all respondents that do not fall into one of 
the extreme sub-groups an increasing income only has a small negative or even a favorably effect on the frequency of doctor's visits. 

In terms of the residual deviance ($4089.9$) and the AIC ($6447.5$) the tree-structured model performs much better than the simple Poisson model with a linear predictor. As it is seen from Figure \ref{fig:trees_visits} the trees are able to capture relations that are hidden by a simple linear predictor.

\section{Concluding Remarks} \label{sec:remarks}

We propose a new tree-based algorithm for the modelling of complex predictor-response relationships using varying coefficients. By recursive partitioning, the method itself identifies the predictors to be modified and the relevant effect modifiers. Main innovations compared to existing approaches are (i) the potential effect modifiers do not have to be specified beforehand, they are automatically chosen from the set of available covariates, and (ii) the linear effect of a covariate is allowed to depend on values of several effect modifiers, in particular on their combination. The visualization of the results as a small tree for each covariate that is modified, enables a simple interpretation of effects and makes it easily accessible to practitioners. 

Although in this article the focus is on generalized linear models with responses from the simple exponential family, the algorithm can easily be extended to more general models. An example are quasi-likelihood models introduced by \citet{Wedderburn:74}, which account for overdispersion in the data. Moreover, the assumption of linear predictors can be weakened, for example, by including polynomial terms. However, one needs to be careful when selecting possible effect modifiers, if variables are present in more than one column in the data matrix.  

All the results of the simulations and applications in this article were obtained by a program that is available from the authors and will soon be made publicly available in an \texttt{R} add-on package.

\bibliography{literatur}

\end{document}